\documentclass [aps,pra,twocolumn,superscriptaddress]{revtex4}
\usepackage{amsmath}
\usepackage{amssymb}
\usepackage{graphicx}
\usepackage[T1]{fontenc}
\usepackage{color}
\newcommand\muvector{\mbox{\boldmath{$\mu$}}}
\newcommand{\comment}[1]{{}}

\begin{document}

\title{On the applicability of quantum-optical concepts in strong-coupling nanophotonics}
\author{C. Tserkezis}
\email{ct@mci.sdu.dk}
\affiliation{Center for Nano Optics, University of Southern Denmark, Campusvej 55, DK-5230 Odense M, Denmark}
\author{A. I. Fern\'{a}ndez-Dom\'{i}nguez}
\affiliation{Departamento de F\'{i}sica Te\'{o}rica de la Materia Condensada and Condensed Matter Physics Center (IFIMAC), Universidad Aut\'{o}noma de Madrid, E-28049 Madrid, Spain}
\author{P. A. D. Gon\c{c}alves}
\affiliation{Center for Nano Optics, University of Southern Denmark, Campusvej 55, DK-5230 Odense M, Denmark}
\affiliation{Center for Nanostructured Graphene, Technical University of Denmark, DK-2800 Kongens~Lyngby, Denmark}
\affiliation{Department of Photonics Engineering, Technical University of Denmark, DK-2800 Kongens~Lyngby, Denmark}
\author{F. Todisco}
\affiliation{Center for Nano Optics, University of Southern Denmark, Campusvej 55, DK-5230 Odense M, Denmark}
%
\author{J. D. Cox}
\affiliation{Center for Nano Optics, University of Southern Denmark, Campusvej 55, DK-5230 Odense M, Denmark}
\affiliation{Danish Institute for Advanced Study, University of Southern Denmark, Campusvej 55, DK-5230 Odense M, Denmark}
\author{K.~Busch}
\affiliation{Max-Born-Institut, Max-Born-Stra{\ss}e 2A, D-12489 Berlin, Germany}
\affiliation{Humboldt-Universit\"{a}t zu Berlin, Institut f\"{u}r Physik, AG Theoretische Optik \& Photonik, Newtonstra{\ss}e 15, D-12489 Berlin, Germany}
\author{N. Stenger}
\affiliation{Center for Nanostructured Graphene, Technical University of Denmark, DK-2800 Kongens~Lyngby, Denmark}
\affiliation{Department of Photonics Engineering, Technical University of Denmark, DK-2800 Kongens~Lyngby, Denmark}
\author{S. I. Bozhevolnyi}
\affiliation{Center for Nano Optics, University of Southern Denmark, Campusvej 55, DK-5230 Odense M, Denmark}
\affiliation{Danish Institute for Advanced Study, University of Southern Denmark, Campusvej 55, DK-5230 Odense M, Denmark}
\author{N. A. Mortensen}
\email{asger@mailaps.org}
\affiliation{Center for Nano Optics, University of Southern Denmark, Campusvej 55, DK-5230 Odense M, Denmark}
\affiliation{Center for Nanostructured Graphene, Technical University of Denmark, DK-2800 Kongens~Lyngby, Denmark}
\affiliation{Danish Institute for Advanced Study, University of Southern Denmark, Campusvej 55, DK-5230 Odense M, Denmark}
\author{C. Wolff}
\email{cwo@mci.sdu.dk}
\affiliation{Center for Nano Optics, University of Southern Denmark, Campusvej 55, DK-5230 Odense M, Denmark}

\begin{abstract}
Rooted in quantum optics and benefiting from its well-established
foundations, strong coupling in nanophotonics has experienced
increasing popularity in recent years. With nanophotonics being
an experiment-driven field, the absence of appropriate theoretical
methods to describe ground-breaking advances has often emerged as
an important issue. To address this problem, the temptation to directly
transfer and extend concepts already available from quantum optics
is strong, even if a rigorous justification is not always available.
In this Review we discuss situations where, in our view, this
strategy has indeed overstepped its bounds. We focus on
exciton--plasmon interactions, and particularly on the idea of
calculating the number of excitons involved in the coupling.
We analyse how,
starting from an unfounded interpretation of the term $N/V$
that appears in theoretical descriptions at different levels of
complexity, one might be tempted to make independent assumptions
for what the number $N$ and the volume $V$ are, and attempt to
calculate them separately.
Such an approach can lead to different, often contradictory
results, depending on the initial assumptions (e.g.
through different treatments of $V$ as the --- ambiguous
in plasmonics --- mode volume). We argue that the source of
such contradictions is the question itself -- \textit{How many
excitons are coupled?}, which disregards the true nature of the
coupled components of the system, has no meaning and often not even
any practical importance. If one is interested in validating the
quantum nature of the system --- which appears to be the motivation
driving the pursuit of strong coupling with small $N$ --- one could
instead focus on quantities
such as the photon emission rate or the second-order correlation
function. While many of the issues discussed here may appear
straightforward to specialists, our target audience is predominantly
newcomers to the field, either students or scientists specialised in
different disciplines. We have thus tried to minimise the occurrence of
proofs and overly-technical details, and instead provide a qualitative
discussion of analyses that should be avoided, hoping to facilitate
further growth of this promising area.
\end{abstract}
\maketitle


\section{Introduction}\label{Sec:intro}

The use of analogies and transference of concepts from other
scientific areas, particularly solid-state physics and quantum
optics, has contributed greatly to the rapid progress
nanophotonics has experienced in recent years. This is
illustrated, for instance, by the development of photonic
crystals~\cite{Joannopoulos_2008}, the quest for Anderson
localisation of light~\cite{segev_natphot7}, or plasmon-induced
transparency~\cite{zhang_prl101}. A more recent example can be
identified in the increasingly popular exploration of
strong coupling in light--matter interactions at the nanoscale.
Starting from the purely quantum Rabi problem for a two-level
system in a dielectric microcavity~\cite{Mandel_1995}, this
area of light--matter interactions has gradually shifted towards
systems exhibiting (semi)classical characteristics, with stronger
(possibly collective) transition dipole moments and modes
confined in ever smaller subwavelength cavities. More than
merely an analogy, this shift of attention has likely occurred
from the necessity to design experimental set-ups that test and
utilise quantum-optical theories
while providing a platform for the realisation of practical
quantum technologies.
The initial study of atoms
placed between dielectric mirrors -- the playground of cavity
quantum electrodynamics (cQED)~\cite{haroche_phystoday42} --
gave thus its place to quantum wells between Bragg
reflectors~\cite{weisbuch_prl69}, quantum dots in photonic
crystals~\cite{yoshie_nat432}, and more recently to excitons
in molecular aggregates or transition-metal dichalcogenides
(TMDs) interacting with plasmonic nanoparticles
(NPs)~\cite{bellessa_prl93,liu_natphot9,liu_acsphot5}.

In the pursuit of stronger emitter--cavity interactions exhibiting
wider avoided crossings between the initial bare states (Rabi
splitting), which will potentially result in stronger quantum
and nonlinear effects~\cite{astafiev_science327,sanvitto_natmat15},
one constant guide has been the factor $N/V$~\cite{torma_rpp78},
where $N$ is the number of emitters, and $V$ a volume to be
discussed, typically of the cavity or the associated optical mode.
In a purely classical description of coupled harmonic oscillators,
this term appears directly as a density of emitters (more generally,
dipole moments)~\cite{torma_rpp78}. On the other hand, from the
standpoint of cQED,
an analysis based on the Tavis--Cummings
Hamiltonian~\cite{tavis_pr170} shows that the collective
coupling strength $g$ of $N$ two-level emitters in a
\emph{low-loss, nonradiative, single-mode} cavity of volume
$V$ is $g \propto \sqrt{N/V}$~\cite{Scully_1997}, where $V$
emerges as a result of the particle-in-a-box quantisation
of the EM fields. The vast majority of experimental and
theoretical efforts in nanophotonics has focused exactly
on enhancing $g$,
either by directly controlling the density of
emitters~\cite{shi_prl112}, or by separately increasing the
numerator or decreasing the denominator of $N/V$. In the
latter approach,
the number of emitters can be precisely controlled when atoms,
quantum dots, or vacancies in solid--state platforms are
involved. This, however, is not necessarily the case when dealing
with the molecular $J-$aggregates that have recently been introduced
in nanophotonics~\cite{dintinger_prb71,sugawara_prl97},
for which one can indeed question the meaning of $N$, as we
discuss below. Current efforts are more devoted on
minimising $V$ through the design of subwavelength
cavities~\cite{chikkaraddy_nat535}, aiming now not only to achieve
as large a $g$ as possible, but do so with the smallest $N$,
so as to potentially enter the quantum regime; this is where
plasmonics comes into play. With these complementary recipes,
extremely high coupling strengths have been achieved, with Rabi
splittings on the order of hundreds of meV (for an overview of recent
experimental results see for example~\cite{baranov_acsphot5}),
even
entering the ultrastrong coupling
regime~\cite{schwartz_prl106,cacciola_acsnano8,todisco_acsphot5}, 
where the rotating wave approximation, which neglects rapidly
oscillating terms,	 no longer holds.

Despite these achievements, however, to quote Hugall et
al.~\cite{hugall_acsphot5}, the efforts of nanophotonics
are \emph{``rarely taken seriously in the context of cQED.''}
Apart from the persistent, ever-present issue of Ohmic
loss~\cite{khurgin_natnano10}, one possible reason for this
could be that in nanophotonics, on some occasions and driven
by the need to shed light on experimental findings, concepts
from other fields have been straightforwardly adapted and
applied outside their original context. Here we discuss some
recent trends that in our view are concerning, focusing
mainly on the attempt to estimate the number of excitons
participating in the collective strong coupling
phenomena~\cite{zheng_nl17,wen_nl17,han_acsphot5,cuadra_nl18}.
We would like to stress that our goal is simply to call
attention to such misconceptions and, hopefully,
contribute thus to the development of the field on more solid
ground. In doing so, we do not intend to undervalue any
previous works --- some of the potential errors that we discuss below
might be found in some of our papers as well~\cite{tserkezis_acsphot5} ---
and our arguments do not invalidate any publication in its
entirety; only small parts of the analysis presented in some works
are criticised.
Evidently, different topics touched upon here might
appear obvious to readers with different backgrounds. This
Review is therefore addressed mostly to newcomers and students,
but if controversy emerges, it should act as an open invitation
to fruitful discussions, which will only help nanophotonics
further advance: cQED already benefited from such debates a
few decades ago~\cite{khitrova_natphys2}. Naturally,
there are several occasions where we agree that strong coupling
was addressed in a careful and productive manner and could act
as guidelines for future research, and some of these are
mentioned in the last section.

\section{Analysis of strong coupling and practices that call for attention}\label{Sec:analysis}

\subsection{The origin of N/V in the coupling}\label{Subsec:origin}

In our experience, the origin of most controversies lies in the
temptation to rely, in every kind of situation, on the readily
accessible but not always applicable equation for the coupling
strength~\cite{bahsoun_acsphoton5,kaluzny_prl51,gao_natphot12}
\footnote{Note that in these references use of
equation~(\ref{Eq:gCoupling}) was fully justified.},
\begin{equation}\label{Eq:gCoupling}
g =
\frac{1}{\hbar} \sqrt{N} 
\muvector \cdot \mathbf{E} =
\mu \sqrt{\frac{\omega N}{2 \hbar \varepsilon \varepsilon_{0} V}}
\propto \sqrt{\frac{N}{V}}~,
\end{equation}
where $\muvector$ is the dipole moment of each one of $N$
identical emitters, $\mathbf{E}$ is the homogeneous electric
field they experience in a good photonic cavity of volume $V$,
$\omega$ the angular frequency of the light, and $\varepsilon_{0}$ and
$\varepsilon$ the vacuum and cavity permittivity, respectively.
This is the result of a fully quantum description, where the
emitters are described as two-level systems and the electromagnetic
(EM) field is quantised into orthonormal modes of a resonator. On
the right-hand side of equation~(\ref{Eq:gCoupling}) we have
deliberately isolated the term $\sqrt{N/V}$, on which we focus
here, because it does not necessarily require a fully quantum
description to emerge. In order to discuss why it is not always
interpreted correctly, it is useful to first briefly consider
the different ways to obtain it, namely a classical,
semiclassical, and quantum description of the problem,
following the analysis previously outlined in \cite{torma_rpp78}.
It is also useful to note at this point that the square-root (rather
than linear) dependence of the mode splitting on $N$ is strong
evidence of the collective response of the ensemble of $N$ individual
emitters, and offers thus an experimentally accessible tool for
verifying the origin of any observed anticrossing.

\begin{figure}[h]
\centering
\includegraphics[width = \columnwidth]{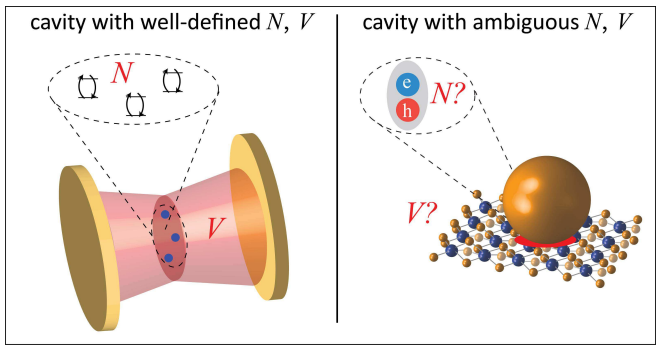}
\caption{Schematic of the physical problem under study. On
the left-hand panel, a collection of $N$ individual
two-level systems is placed between two mirrors forming
a dielectric microcavity of volume $V$. Both $N$ and $V$
are well-defined quantities, and using the Tavis--Cummings
picture makes sense. On the right-hand panel, excitons in
a TMD interact with a plasmon mode of a metallic NP.
Both $N$ and $V$ are ambiguous, and usage of the $N/V$
factor is questionable.}\label{fig1}
\end{figure}

With strong coupling in nanophotonics involving large
aggregates of organic molecules or layered TMDs (where the
dynamics of excitons is strongly influenced by long-range
Coulomb interactions) combined with extended metallic films or small
(but macroscopic) plasmonic NPs, a purely classical description
can already capture most of the observed effects. In this case,
one can start, as a first approximation, by considering a single
electron of the emitter layer as a damped harmonic oscillator, driven
by the external --- homogeneous and harmonic --- EM field, with its
dipole moment acquiring the form of a classical Lorentzian 
oscillator. In a macroscopic medium with $N$ dipoles
per unit volume $V$, the polarisation density $\mathbf{P}$
is defined as the average dipole moment per unit volume
\cite{Jackson_1999}, from which the resulting susceptibility
and permittivity of the emitter layer depend on $N/V$.
In the simple example of a dye layer coupled to the surface
plasmons of an underlying metal film, one can introduce the
dye permittivity in the plasmon dispersion, leading to a
quadratic equation with two branches separated by a band
gap whose width is proportional to $\sqrt{N/V}$. A larger effective
\emph{density} of dipoles provides thus the straightforward
means to obtain wider stop bands, i.e. mode splittings, as
has been shown experimentally~\cite{hakala_prl103}, and this
density is the only relevant parameter.

In a semiclassical description, the EM field remains
classical but the homogeneous emitter layer is replaced
by a collection of two-level quantum emitters. Starting
with the case of a single emitter, the Hamiltonian of the
system, within the rotating-wave approximation and after
the appropriate unitary transformations, is written as
\begin{equation}\label{Eq:RWAHamiltonian}
H =
-\frac{\hbar \delta}{2} +
\frac{\hbar \Omega_{\mathrm{R}}}{2}
\left(\sigma_{+} + \sigma_{-}\right)~,
\end{equation}
where $\Omega_{\mathrm{R}}$ is the Rabi frequency, $\delta =
\omega - \omega_{0}$ is the detuning between the frequency
of the optical mode $\omega$ (say the surface plasmon,
to relate with the classical description) and the frequency
$\omega_{0}$ of a transition from the excited to the ground
state of the emitter, and $\sigma_{\pm}$ are the standard
Pauli matrices~\cite{Walls_1994}. Making the extension to
a collection of such emitters, and calculating the expectation
value of the induced dipole moment, it can be shown (for
details see \cite{torma_rpp78}) that the polarisation density
for $N$ emitters per unit volume is again related to the
density $N/V$, with $\Omega_{\mathrm{R}}\; (= 2g)\; \propto
\sqrt{N/V}$. When $N$ emitters oscillate in phase, the
obtained Rabi splitting is proportional to the emitter
density~\footnote{It should be noted here that both classical and semiclassical
descriptions have been obtained within certain approximations,
and deviations from the $\sqrt{N/V}$ dependence might be observed
for small numbers of emitters in small volumes --- when a
macroscopic density is not well defined. Nevertheless,
this dependence has been observed so frequently in
experiments, that any microscopic model is expected
to lead to a similar response.}.

It is only in a fully quantum description, where the
EM field is also quantised and both the emitter and
the field are described by appropriate creation and
annihilation operators, that the two factors, $N$ and $V$,
can be rigorously defined separately. As implied by
equation~(\ref{Eq:gCoupling}), $V$ is in this case directly
related to the EM field and, in particular, to its
particle-in-a-box normalisation. In this case, the system
is described by the Tavis--Cummings Hamiltonian, which is
often solved for simplicity in the large-$N$ limit, with
only a small fraction of the total emitters being excited.
This allows to map the many two-level (fermionic) emitters
to a large bosonic super-oscillator, to obtain
$\Omega_{\mathrm{R}} = 2g$ with $g$ given by
equation~(\ref{Eq:gCoupling}).
For $N =1$, the first equality in
equation~(\ref{Eq:gCoupling}) is thus an accurate
starting point, as it describes light--matter interactions
within the dipole approximation~\cite{Walls_1994,Agarwal_2012},
although one could, and in some situations should, go one
step backwards, e.g. by introducing the electron--photon
interaction Hamiltonian in the appropriate gauge to ensure
that higher-order interaction terms are
included~\cite{kurman_natphot12,cuartero_acsphon5}. Using
the same equations for $N > 1$ yet distinguishable emitters,
however, requires that all dipole moments experience
a \emph{uniform} electric field of the cavity and (for the
second equality) they are \emph{aligned}. For atoms in extended
microcavities excited by a laser field, a description in terms
of two-level systems with the same dipole moment experiencing
the same field is valid, and an analysis based on
equation~(\ref{Eq:gCoupling}) can safely
apply~\cite{rempe_prl58,thompson_prl68}. In contrast, in
plasmonics, where the cavities are i) open, ii) comparable in
size with the emitters themselves, and iii) lossy, such
an approach must be judiciously adopted, as recent
experiments corroborate~\cite{yankovich_nl19}. Below we
discuss some of the ambiguities that may arise in defining $V$
and $N$, shown schematically in Figure~\ref{fig1}, as they
appear in recent literature.

\subsection{Mode volume: is it always the same?}\label{Subsec:volume}

The mode volume enters equation~(\ref{Eq:gCoupling}) as the
normalisation constant for the quantised field, denoting
the volume of a hypothetical cavity that would provide the
same maximum field per photon~\cite{gerard_jlt17}. But this
already requires that the field \emph{can} be quantised in an
unambiguous manner. Whether and how one can define a mode volume
for an open, dispersive cavity with high Ohmic and radiative
losses is still under debate. The simple relation based on
energy density~\cite{Novotny_2006} --- often suitable for
dielectric cavities --- no longer holds, and a first required
action is to at least account for the dispersive character of
the metal~\cite{ruppin_pla299,maier_oex14}. In such an approach
one usually normalises a volume integral of the energy density
to its maximum value. But particular care must be taken here:
such a normalisation might be an acceptable starting point to
describe the plasmonic mode itself, but is not necessarily
extendable to the emitter--cavity problem, as we discuss below.
More rigorous definitions and derivations of mode volumes
have been obtained in recent years in the context of quasinormal
modes (QNMs)~\cite{sauvan_prl110,kristensen_ol37,kamandar_prb97},
obtained as the solutions $\tilde{\mathbf{f}}_{\mu}$ of
the Helmholtz equation with open boundary conditions,
\begin{equation}
\boldmath{\nabla} \times \boldmath{\nabla} \times
\tilde{\mathbf{f}}_{\nu} (\mathbf{r}) -
\left( \frac{\tilde{\omega}_{\nu}}{c} \right)^{2}
\varepsilon (\mathbf{r}, \omega)
\tilde{\mathbf{f}}_{\nu} (\mathbf{r})
= 0~,
\end{equation}
where $\tilde{\omega}_{\nu}$ is the complex resonance frequency
of mode $\nu$, and $\varepsilon (\mathbf{r}, \omega)$ the position
($\mathbf{r}$)- and frequency-dependent permittivity of the cavity,
a formalism which even allows to straightforwardly implement
quantum corrections~\cite{kamandar_optica4}. Once these solutions
are normalised, one can expand any field in a set of QNMs, construct
the Green's function, and in principle obtain the solution to any EM
problem. But the exact normalisation is still under debate: it is for
instance
an open question whether it is preferable to base
a QNM formulation on a Green's function
expansion~\cite{leung_pra49,kristensen_pra92} or a field
expansion~\cite{sauvan_prl110,yan_prb97,lalanne_josaa36}.
While we do not have a clear answer to this, we believe that QNMs,
or possibly hybrid schemes with elements from both a quasi- and
a pseudo-mode picture~\cite{dalton_pra64},
even if not always practical for an immediate analysis~\cite{yang_nat576},
constitute a promising prospect. For more details about different approaches
to QNMs we refer the reader to the Perspective by Kristensen
and Hughes~\cite{kristensen_acsphoton1}, the Review by Lalanne
\emph{et al.}~\cite{lalanne_lpr12},
or the more recent Tutorial by Kristensen
\emph{et al.}~\cite{kristensen_arxiv1910}.

In the context of emitter--cavity coupling, a frequent approach
to obtain a first estimate for the volume that enters
equation~(\ref{Eq:gCoupling}) is based on simple geometric
arguments. For example, what is sometimes used as $V$ is the
geometric volume of the plasmonic NP~\cite{zengin_prl114}.
The reasoning behind this choice is based on the small
dimensions of the NP (thus less radiative), and the fact
that the energy is confined inside it. While this may
offer a rough qualitative estimate for the plasmonic mode
volume itself --- within an order of magnitude as compared
e.g. to QNM approaches --- this approach is not
particularly accurate for emitter--plasmon coupling. Here,
the volume that matters is only that which is relevant to
the emitter, and appears in calculations based on the local
density of states (LDOS)~\cite{koenderink_ol35,shahbazyan_prb98}
through the appropriate Green's function~\cite{delga_prl112}.
It is clearly mentioned for instance by
Shahbazyan~\cite{shahbazyan_nl19} that with such robust
calculations one can only obtain a \emph{local} mode
volume. And since volume integrals always appear in the
relevant expressions, it is crucial to distinguish between
integrating over the entire space or over the entire
\emph{cavity}. In such situations, single NPs, or their
assemblies, do not constitute the cavity, but resemble more
a combination of the exciting source and the mirror at the
cavity end. On the one hand, it is localised plasmon
resonances that generate the enhanced near field where the
emitters are placed, boosting thereby the LDOS (source part).
On the other hand, these are also the origins of loss, through
radiation and absorption, so that in this respect they are
(bad) cavity mirrors. Eventually, what is really relevant is
the electric field confined near or between NPs (and/or a possible
substrate), where the emitters are located~\cite{ramezani_prb94}.
For rough estimates of orders of magnitude, a better --- though
still loose --
approximation would be thus to just consider the geometric
volume of the cavity, e.g. the volume between two
spheres in a plasmonic dimer~\cite{roller_nl16}.

Finally, another mode-volume calculation that occasionally appears
in literature, is the one related to the Purcell
factor~\cite{purcell_pr69},
\begin{equation}\label{Eq:Purcell}
P =
\frac{3 Q}{4 \pi^{2} V}
\left( \frac{\lambda}{n} \right)^{3}~,
\end{equation}
where $n = \sqrt{\varepsilon}$ is the refractive index and
$Q$ the quality factor of the cavity. Since $P$ can be
evaluated (typically with numerical calculations) from the
shortening of the excitation lifetime, and $Q$ is given
by the resonance linewidth, obtaining $V$ from
equation~(\ref{Eq:Purcell}) appears tempting, and a Letter
by Koenderink~\cite{koenderink_ol35} is often cited as the
reference which justifies its use. However, as clearly
stated in that Letter, this equation holds only as long
as \emph{a)} normal modes \emph{can} be defined for the
system, so that the LDOS is written as a sum of such, and
\emph{b)} this sum is dominated by a single
mode. This is usually not the case in plasmonics, especially
as far as the single-mode requirement is concerned, and
the shortening of the lifetime is often different (larger)
than the emission intensity enhancement due to the coupling
to both radiative and nonradiative plasmonic modes.
Nevertheless, equation~(\ref{Eq:Purcell}) is still in use
--- although usually not as a means to obtain quantitative
results~\cite{chikkaraddy_nat535} --- even though the very
occurrence of strong coupling and the suppression of the
well-known fluorescence quenching~\cite{anger_prl96} occurs
exactly because the emitters also interact with a
pseudomode formed by all higher-order nonradiative
modes~\cite{delga_prl112,ruppin_jcp76,kongsuwan_acsphot5}. We
could not help noticing that situations of using the equations
in \cite{koenderink_ol35} out of context must be so frequent,
that its author felt the need to clarify in a recent
paper that \emph{``we use the term mode volume here not as an
endorsement of the validity of this concept per se for
plasmonics''}~\cite{palstra_nanophot8}.

To summarise this subsection, the main idea we try to
convey is that the appearance of a $V$ in an equation
does not necessarily always mean \emph{the same} volume:
different effects should require a different treatment.
In this respect, the volume that enters normalisation
of the plasmonic field, calculated either through the
energy density or QNMs, is not the same as the one that
is relevant to an emitter, which is related to the LDOS the
emitter experiences at points \emph{outside} the
cavity. The concept of different mode volumes is
well-established, for example in the context of plasmonic
waveguides~\cite{oulton_njp10}.

\subsection{Number of emitters: should one count excitons?}\label{Subsec:number}

The previous subsection discussed different ways to
estimate a mode volume, and the ambiguities that
emerge in the case of plasmonics. But even if it
would eventually be calculated with a universally accepted
method, the question that lies at the heart of this
Review is whether that volume should in turn be used to
evaluate the number of emitters coupled to a plasmonic cavity
from the observed mode splitting.
In 2016, Chikkaraddy \emph{et al.}~\cite{chikkaraddy_nat535}
studied a carefully designed cavity where individual molecular
emitters, encapsulated in barrel-shaped molecules so as to
prevent them from interacting with each other and align their
dipole moments with the plasmonic near field, were placed in
ultrasmall NP-on-mirror cavities. Statistical analysis of the
data showed that, in some situations, the occurring anticrossing could be attributed
to interaction of the cavity with just one emitter, thus
reporting single-molecule strong coupling at room temperature.
This
result, together with the work published on the same day by
Santhosh \emph{et al.}~\cite{santhosh_natcom7}, involving
a single quantum dot in the middle of a plasmonic bowtie antenna,
has set strong-coupling nanophotonics on a new basis.
But in the wake of the enthusiasm that followed those papers, a number
of other works seem to have initiated --- albeit unintentionally --- an
undeclared
competition for the strongest coupling with the smallest number
of emitters (presumably with the intention to enter the quantum-optical
regime), which is still growing today. Even though the conclusions of
those seminal experiments were not based on
equation~(\ref{Eq:gCoupling}), the $N/V$ factor has turned
into the most easily accessible analysis tool in recent
literature, either merely for qualitative discussions (where it
\emph{can} be legitimate) or even to obtain quantitative
results~\cite{todisco_acsphot5,han_acsphot5,tserkezis_acsphot5,
gao_natphot12,kewes_acsphotn5}. In the latter case, $g$ is usually
first estimated from the experimentally or numerically
observed spectral anticrossing as $g = \Omega_{\mathrm{R}}/2$
(assuming a lossless system). If the values of both $N$ and
$V$ are well-defined and thus meaningful, the coupling strength
obtained from their ratio is then compared to the observed
$g$ to check the validity of equation~(\ref{Eq:gCoupling})
and verify that the strong-coupling regime has been reached.
This control can of course also be done when the
density $N/V$ is directly available.
But when collective excitations are involved and $N$ is
ill-defined, researchers have nevertheless been tempted to use $g$
and the $V$ obtained either from rough geometrical
estimates~\cite{roller_nl16,stete_jpcc122}, or numerically
as described in the previous subsection, to evaluate $N$.
Whether such a calculation is meaningful, is the topic
of this subsection.

Ever since excitons in $J$-aggregates of organic molecules
were first proposed as systems with a higher dipole moment,
which can thus facilitate strong coupling in
plasmonics~\cite{bellessa_prl93,pockrand_jcp77}, several
attempts to extend the use of equation~(\ref{Eq:gCoupling})
to any strong-coupling set-up, regardless of the nature of
the emitter, have been made. In particular, excitons in
TMDs coupled to plasmonic NPs have emerged as an attractive
architecture~\cite{zheng_nl17,wen_nl17,cuadra_nl18}; yet
in the case of TMDs
--- whose complex excitonic resonances are described beyond
a single-particle picture~\cite{wang_rmp90} ---
this approach has to be questioned: this
is really more a problem of interpretation than of the
calculation itself. It is of course always possible to
calculate a value $N$ from a coupling constant $g$, a
mode volume $V$ and a permittivity $\varepsilon$ (whatever
the appropriate one might be for a plasmonic resonator on
a substrate in air), but it is not clear what this --- in
general non-integer --- $N$ might be counting. As we 
show further below, it certainly does \emph{not} count
the number of excitons that are coupled to the plasmonic
resonator in the way isolated molecules do. 

\begin{figure}[h]
\centering
\includegraphics[width = \columnwidth]{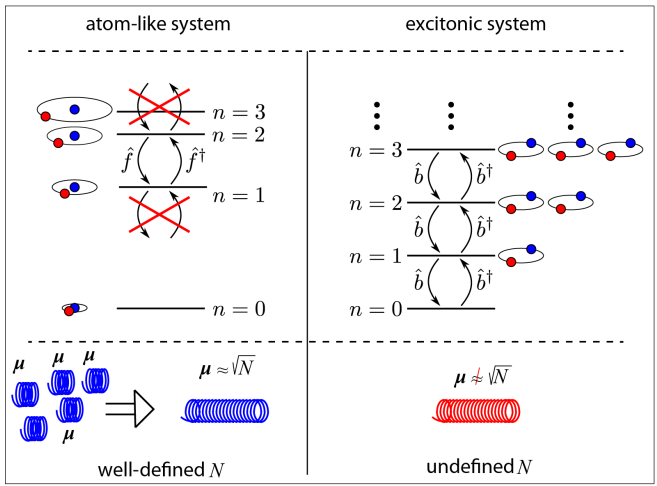}
\caption{Differences between the types of emitters
employed in strong-coupling schemes in nanophotonics.
Upper panels: On the left, the energy ladder of a two-level
system; the presence of $N$ emitters (e.g. atoms) in the
cavity means exactly $N$ such two-level systems (see
schematics of Figure~\ref{fig1}) interacting with the
cavity field (and possible with each other, so-called
Dicke superradiance). On the right, the energy ladder
of a bosonic system (e.g. excitons in TMDs). Adding one
exciton means creating one additional boson and
going one step up the ladder, with all excitons sharing the
same quantum numbers and contributing to the collective
excitonic state that interacts with the cavity. Interactions
between excitons and formation of more complex quasiparticles
like bi-excitons or trions have been  disregarded for the
sake of simplicity.
Lower panels: On the left, $N$ independent two-level systems
with dipole moment $\muvector$ combine in the low-excitation
limit to form a super-oscillator whose dipole moment
is proportional to $\sqrt{N}$. On the right, the strong
collective dipole moment of an excitonic system is not the
result of the combination of $N$ specific dipoles.
}\label{fig2}
\end{figure}

Our main argument against this practice is based on the
apparent mix-up between a physical system (as described,
e.g., by the Hamiltonian) and its states. Single molecules
(or atoms) are clearly part of the system, represented by
a pair of raising and lowering operators $\hat{f}^{\dag}$
and $\hat{f}$ per molecule in the Hamiltonian. In contrast,
an exciton is \emph{not} part of the bare system, and the excitonic
population does not appear directly or indirectly anywhere
in the Hamiltonian. An exciton is an excitation of the
system --- hence the name. The excitonic population is an
observable, associated with one of several quantum numbers
that can be used to label the state of the system. When
compared to a single quantum emitter like an atom or colour
center, the exciton does not correspond to an emitter, but
rather to a \emph{transition} within an
effective emitter. A schematic representation of the
differences between atom-like and excitonic emitters is
shown in Figure~\ref{fig2}. In the upper left-hand panel,
the energy ladder of a single two-level system is shown.
For $N$ emitters one has to simply imagine $N$ identical
such ladders. Instead, in TMDs there exists an extended
distribution of excitons --- and the excitonic gas is
bosonic (or at least nearly
so)~\cite{hopfield_pr112,combescot_epl59,laikhtman_jpcm19}.
While the classification of excitons in TMDs is not perfectly clear,
the fact that their wavefunctions have the form of
Wannier--Mott excitons~\cite{yu_nsr2,cudazzo_prl116},
implies that they should definitely not be treated as
two-level systems --- even though their binding energies
are closer to those of the more molecule-like Frenkel
excitons. 
Bosonic creation and annihilation operators $\hat{b}^{\dag}$
and $\hat{b}$ generate a new electron-hole pair going one step
up the bosonic ladder, as shown in the figure within a rather
simplified picture where exciton interactions and formation
of bi-excitons~\cite{mak_natphot10,he_natcom7} are disregarded
for simplicity \footnote{
Here it is important to distinguish excitons in, say, quantum dots
--- where any electron--hole pair is (or, rather, \emph{becomes}) an exciton ---
and in TMDs, where there is a crucial difference between an excited
electron--hole pair (single-particle picture, no interactions) and an
exciton (many electron picture, screened interaction).}.
It is therefore important to depart from the
picture of excitons being analogous to a hydrogen atom,
and distinguish between physical entities and quasiparticles.

One could argue that the fundamental issue regarding the nature
of the coupled components
can be bypassed by the transformation of $N$ two-level
systems to a bosonic super-oscillator, as discussed in
Subsection~\ref{Subsec:origin} and shown in the bottom
schematics of Figure~\ref{fig2}. This is valid, provided
that the system is in the linear response regime
(low number of excitations so that saturation effects are
irrelevant, which is implicitly assumed throughout this
manuscript). But apart from this,
several practical problems arise in such calculations. As
discussed above, the first one appears in the form of the
definition of the cavity: the dipole moment of TMDs is strictly
restricted in the two-dimensional material layer, and aligned in
plane. Evaluating the energy density everywhere in space around
a plasmonic NP on top of such a TMD~\cite{hou_aom7} is thus
really not relevant. Furthermore, in estimations based on $N/V$
one typically assumes that the coupling strength is that
corresponding to an emitter located at the position of the
\emph{maximum field}, which is definitely inaccurate for a large
collection of excitations distributed over a large area and
inside the inhomogeneous near field of a plasmonic antenna
of any shape. The dipole moment in such a calculation can also
be ambiguous, and it is tempting to use equation~(\ref{Eq:gCoupling})
as its indirect measure~\cite{stuhrenberg_nl18}.
One could raise the question about the nature of the matter
excitations the cavity couples to: are they individual,
site-localised excitons, or an excitonic state extended in
space and involving a number of bound electron--hole pairs in the
TMD sheet? As already pointed out by others~\cite{stuhrenberg_nl18},
only the latter can account for the strong dipole moment
achieved in such systems. While this seems again to be the case
of a super-oscillator, similar to that of $N$ two-level atoms
treated by the Tavis--Cummings Hamiltonian, the key difference
is that excitons are delocalised or, at best, only localised over
many unit-cells in the TMD and do
not correspond to a specific number $N$, as shown in the bottom
schematics of Figure~\ref{fig2}, whereas the dipole moment
of the molecular super-oscillator is indeed proportional to $\sqrt{N}$.
This super-oscillator response
was the charm of $J$-aggregates in the first place, with
dipole moments aligning to create the stronger effective
dipole responsible for the anticrossing in the spectra.
But then it is clear that a calculation of a number of
excitons uniquely on the basis of the dipole moment of \emph{a single}
exciton is misleading and highly questionable. One should thus
be particularly careful not to mix extended hybrid states
with transitions in distinguishable entities.

Such ambiguities have led to estimations of very different
numbers of excitons for very similar systems, most
strikingly in the case of metallic nanorods on top of
tungsten-based TMDs, where in \cite{wen_nl17}
a number of about $5$ excitons was estimated, while in
\cite{zheng_nl17} the corresponding
value was of the order of a few thousands. Despite the
obvious inconsistency, the apparent appeal of such calculations
persists, and they are now adopted even in other contexts
where no emitters are present, to evaluate for instance the
dipole moment of individual plasmons in metallic nanodisc
arrays, where each disc is essentially treated as the equivalent
of a two-level system~\cite{bisht_nl19}, using the standard
terminology of the Rabi problem. The problems with this
approach are that
\emph{a)} not every hybridisation gap can be called a
Rabi splitting, neither can every excitation (in this
case the plasmon) be assimilated to an idealised (two-level) system
with a transition. This might seem like a matter of
terminology, but it is exactly the kind of approach that
can lead to the unfounded extensions of concepts discussed
in this Review. \emph{b)} In any case, the disc array supports
a collective plasmonic mode~\cite{tserkezis_jpcm20} and
conclusions about individual dipole moments cannot be judiciously
drawn.
While estimating a dubious exciton number does
not provide any particularly useful information in this
context of strong coupling (at least in the majority of scenarios),
if one insists, there are still more accurate ways than
through $N/V$ to do this. An example of a more appropriate
approach can by found in~\cite{zheng_nl17} where, starting from the
first equality in equation~(\ref{Eq:gCoupling}), the authors
evaluated a \emph{local} coupling,
at position $\mathbf{r}_{i}$, through $\muvector
(\mathbf{r}_{i}) \cdot \mathbf{E} (\mathbf{r}_{i})$ and
summing over all local $g$ values:
\begin{equation}\label{Eq:CouplingFromDipole}
g = \sqrt{\sum_{i} g (\mathbf{r}_{i})^{2} }~.
\end{equation}
Since the EM field decays exponentially away from the plasmonic
nanostructure, the sum in equation~(\ref{Eq:CouplingFromDipole})
will eventually converge to some value, which in \cite{zheng_nl17}
agreed rather well with a rough estimate based on the single
exciton radius and the area below the metallic nanorod. Such
a calculation could be useful in describing the system in terms
of the equivalent image of individual excitons that \emph{would}
produce the same response --- or, in other words, an \emph{effective}
exciton. Nevertheless, even this approach contains the somehow vague
notation $i$: a space discretisation immediately reduces the problem
to the traditional picture of $N$ two-level systems, and if all their
dipole moments are assumed equal, aligned, and experiencing the
same field, one retrieves equation~(\ref{Eq:gCoupling}). But to use
this for a very rough estimate, one can only base it on the assumption
that each exciton occupies a specific space in the TMD sheet
(typically defined by its Bohr radius), an assumption in
strong conflict with the collective character of the
excitation. A more strict calculation should include an
integral of the coupling strength over the entire
TMD sheet area $\mathcal{S}$~\cite{schlather_nl13}
\begin{equation}\label{Eq:2DsheetIntegral}
\int_{\mathcal{S}} \mathrm{d}^{2} r \; |g(\mathbf{r})|^2~,
\end{equation}
as we briefly show with a simple toy model in the next
Subsection. But the main message of this part is that one
should not try to extend the analogies between excitonic
states and two-level systems beyond certain limits.

\subsection{Size of excitons: what is the coupling strength?}\label{Sec:bosons}

In the previous Subsection we pointed out that excitons,
for example in atomically-thin TMDs, should not be confused with atoms,
as their number is not an inherent property of the system
itself (i.e., the Hamiltonian), but a consequence of the
illumination
--- and it should be clear that all the treatments discussed
here only apply in the low excitation, linear-response case
anyway.
A second misunderstanding that can easily develop from
the exciton--atom analogy is that, despite the illustrative
picture of a single exciton as a bound state of orbiting
electron and hole (Figure~\ref{fig2}), the excitonic state
in a TMD coupled to a plasmonic resonator is in fact
\emph{not} localised at the excitonic Bohr radius
--- one of the reasons being exactly the fact that the
plasmon--exciton interaction gives rise to new, hybrid states,
which are evidently dissimilar to the initial bare states.
This implies that the effective plasmon-exciton coupling cannot
be estimated via the \emph{maximal} field enhancement of the
plasmonic structure, a treatment based on the picture of
excitons as small movable atom-like objects that accumulate
in the plasmonic hotspot and all couple according to the
maximally attainable electric field. This treatment is
inaccurate, at least within the regime where the optical
response of the TMD is approximately linear, i.e. for
low to moderate optical excitation intensities. To illustrate
this, we introduce a very simple model, where we bypass
the mode volume issue by assuming a lossless closed
cavity. Although our description is highly idealised and
in principle well-known~\cite{dhara_natphys14,flick_nanoph7},
it is capable of explicitly showing how extended excitonic
states emerge quite naturally even from a model based on
non-interacting and localised excitons. While there are
analogies with the Dicke~\cite{dicke_pr93} or
Tavis--Cummings~\cite{tavis_pr170} models for two-level
emitters, these are not identical cases\footnote{And even if they
were, a direct transition from the Tavis--Cummings to the
Jaynes--Cummings description is not always
straightforward~\cite{trivedi_prl122}.}.

We consider the interaction of an optical mode described
by ladder operators $\hat a$ and $\hat{a}^\dagger$ with
localised exciton states that can be created and annihilated
at any point $\mathbf{r}$ within the TMD via the operators
$\hat{b}^\dagger_{\mathbf{r}}$ and $\hat b_{\mathbf{r}}$,
respectively. The state of an exciton centred at $\mathbf{r}$
is denoted as $| \mathbf{r} \rangle$, while the single-photon
state of the resonator as $| a \rangle$. We further assume
that \emph{a)} both the closed cavity and the excitonic layer
support normalisable modes without damping, and \emph{b)} there
is no interaction between the excitons. The latter is just a
reformulation of our assumption of moderate pumping
intensities, whereas the former is a somewhat crude assumption
for the sake of simplicity. In closed cavities losses can
be introduced via Lindblad operators~\cite{neuman_optica5}.
In open cavities, and/or in the presence of dissipation,
things become even more complicated, but steps towards an
accurate description have been taken recently~\cite{franke_prl122}.
A more precise analysis where interactions between excitons
are taken into account is feasible~\cite{agranovich_prb67},
but it exceeds the purely illustrative purposes of this
section.

Within the rotating wave approximation, the exciton--cavity
Hamiltonian is
\begin{equation}\label{Eq:Hamiltonian}
\mathcal{H} =
\hbar \omega_{0} \hat{a}^{\dagger} \hat{a} +
\int_{\mathcal{S}} \mathrm{d}^{2} r \;
\left\{ \hbar \Omega
\hat{b}_{\mathbf{r}}^{\dagger} \hat{b}_{\mathbf{r}} +
[g (\mathbf{r}) \hat{a}^{\dagger} \hat{b}_{\mathbf{r}} +
\mathrm{h.c.}] \right\}~,
\end{equation}
where $\omega_{0}$ is the angular frequency of the cavity
mode, $\Omega$ the angular frequency of the degenerate
excitons, and $\mathrm{h.c.}$ denotes the Hermitian conjugate
of the expression in square brackets. An exciton at
$\mathbf{r}$ couples to the cavity mode via a matrix element
$g(\mathbf{r})$, whose exact form is of no concern in this
context. In the case of distinguishable emitters it is
usually sufficient to employ the electric dipole
approximation as $g(\mathbf{r}) =  \muvector \cdot
\mathbf{E}(\mathbf{r})$, where $\muvector$ denotes the
exciton dipole moment and $\mathbf{E}(\mathbf{r})$
is the electric field of the optical mode. In order to
diagonalise this Hamiltonian, we introduce a family
$h_{\alpha} (\mathbf{r})$ of functions that are
orthogonal to $g(\mathbf{r})$ and such that the
set $\{ g(\mathbf{r}), h_{\mathbb{R}} (\mathbf{r})\}$
constitutes a basis for the space $L^{2}(\mathcal{S})$.
This should always be possible if the cavity mode is
square-integrable in the sheet.

We now transform the excitonic system into this new
basis $|g \rangle = N_{g} \int_{\mathcal{S}}
\mathrm{d}^{2} r \; g{^\ast} (\mathbf{r})
|\mathbf{r} \rangle,$ and $| \alpha \rangle =
N_{\alpha} h_{\alpha}^{\ast}(\mathbf{r})
|\mathbf{r} \rangle$, where the normalisation
constants are $N_{g} = \left[ \int_{\mathcal{S}}
\mathrm{d}^{2} r \; |g(\mathbf{r})|^{2} \right]^{-1/2}$,
$N_{\alpha} = \left[ \int_{\mathcal{S}} \mathrm{d}^{2} r \;
|h_{\alpha} (\mathbf{r})|^{2} \right]^{-1/2}$.
By applying the Hamiltonian to the orthogonalised states
$| \alpha \rangle$, and using $\hat{a} | \mathbf{r} \rangle = 0$
and $\hat{b}_{\mathbf{r}} | \mathbf{r}' \rangle = \delta(\mathbf{r}
- \mathbf{r}') | 0 \rangle$, we find
$\mathcal{H} | \alpha \rangle = \hbar \Omega | \alpha \rangle$.
Obviously these states do not couple to the resonator mode,
because their spatial envelope was constructed to be orthogonal
to the coupling distribution $g(\mathbf{r})$, and could be
called cavity-dark exciton states. As a result, the light-matter
coupling problem reduces to the two remaining states
$| a \rangle$ and $| g \rangle$. Applying the Hamiltonian
we obtain
$\mathcal{H} | a \rangle = \hbar \omega_{0} | a \rangle +
N_{g}^{-1} | g \rangle$
and
$\mathcal{H} | g \rangle = \hbar \Omega | g \rangle +
N_{g}^{-1} | a \rangle$.
This means that the eigenstates can be written as
$c_{a} | a \rangle + c_{g} |g \rangle$, where the
coefficients are given by the eigenvalue
problem
\begin{eqnarray}\label{Eq:matrix}
\hbar
\left(
\begin{array}{cc}
\omega_{0} & \kappa \\
\kappa & \Omega \\
\end{array}
\right)
\;
\left(
\begin{array}{c}
c_{a} \\
c_{g}
\end{array}
\right)
= 
\mathcal{E}
\left(
\begin{array}{c}
c_{a} \\
c_{g}
\end{array}
\right) ,
\end{eqnarray}
with the real-valued $\kappa = (\hbar N_{g})^{-1}$. The
eigenenergies are
\begin{eqnarray}\label{Eq:eigenenergies}
\mathcal{E}_{\pm} = 
\hbar \frac{\omega_{0} + \Omega}{2} \pm 
\hbar \frac{\sqrt{(\omega_{0} - \Omega)^{2} +
\kappa^{2}}}{2} = 
\hbar \Omega \pm \frac{\hbar \Omega_{\mathrm{B}}}{2} ~,
\end{eqnarray}
where the second equality holds on resonance, with
coefficients $c_{a} = \pm c_{g} = \sqrt{1/2}$ and beat
frequency
\begin{equation}\label{Eq:beatfrequency}
\hbar \Omega_{\mathrm{B}} =
\sqrt{ \int_{\mathcal{S}} \mathrm{d}^{2} r \;
|g(\mathbf{r})|^{2}}~,
\end{equation}
which contains the expression suggested in
equation~(\ref{Eq:2DsheetIntegral}). Similar analyses
were presented recently in \cite{shahbazyan_nl19}
and \cite{franke_prl122}. The corresponding problem
for $N$ two-level quantum emitters in a
plasmonic cavity (where the cavity was indeed such,
consisting of metal films), was solved in~\cite{gonzalez_prl110},
while a computationally efficient model that allows to
describe a large number of two-level emitters in a
microcavity was introduced in~\cite{trivedi_prl122}.
While all these approaches might share some common points,
the key difference is the origin and meaning of the
collective dipole moments involved in the coupling.

The main message from the present analysis is that the eigenstates
of the Hamiltonian are hybrid states that have both plasmonic
and excitonic character, and the excitonic part itself is a
hybridisation of many excitons delocalised throughout the entire TMD
sheet. This is not exactly surprising, as it corresponds to two
harmonic oscillators coupled to each other, where the eigenstates
emerge as the hybridised states, and the linearity of both subsystems
guarantees that these new states are excited quantum-by-quantum.
There is not much sense in counting the number $N$ of excitons,
because their number is strictly linked to the photon number
--- and thus dependent on the illumination intensity ---
via the hybrid eigenstate. 

\section{Discussion and conclusions}\label{Sec:discussion}

The preceding analysis does not imply that
strong-coupling nanophotonics went astray in its
entirety and should be completely revisited: on the
contrary, this is a very fruitful area of research,
where many important results have been derived and
exciting applications suggested~\cite{fernandez_acsphot5}.
For instance,
quantum statistical phenomena such as Bose--Einstein
condensation and polariton lasing do not necessitate operation
at the single-emitter or single-photon limit, on the contrary, they 
require a large number of excitations~\cite{amo_nat457,
daskalakis_natmat13,hakala_natphys14,kena-cohen_natphot4,
ballarini_natcom4,ramezani_optica4,vakevainen_arxiv1905}.
Similarly, nonlinearities and entanglement~\cite{raimond_rmp73}
can be explored without the need for single excitons. In
any case, if one wants to maximise the coupling, classical
physics already provides the guidelines: strong modes (EM
fields) with significant overlap are the safest way to
increase the interaction in any system effectively described
as a pair of coupled harmonic oscillators. Furthermore, for
quantum applications, what one really needs in practice is
single photons~\cite{lodahl_rmp87,pelton_natphot9},
regardless of the way they where generated,
and the questionable single excitons in organic molecules or
TMDs are not, by default, a necessary condition to achieve
this goal. Consequently, it makes much more sense to focus
for example on whether antibunching is observed, and try to
measure quantities such as the photon emission rate and the
second-order correlation function, as was done very recently
for other plasmon--emitter systems~\cite{saez_optica4,
singh_nl18,ojambati_natcom10}.

Another exciting direction that is drawing significant
attention recently is polaritonic chemistry~\cite{feist_acsphoton5},
which focuses on the possibility to manipulate chemical
reactions and structures through the formation of polaritons.
Initiated by the seminal work of the Ebbesen
group~\cite{hutchison_angchem51}, this area is now shifting
from the usual dipole approximations to consider all internal
degrees of freedom (electronic, vibrational, nuclear) of the
molecules~\cite{galego_prx5}, calling thus for a rigorous
theoretical description, possibly within time-dependent
density-functional theory~\cite{neuman_nl18,rossi_natcom10,
flick_nanoph7}, where many of the assumptions criticised here
are absent by construction.

Throughout this manuscript, whenever discussing a treatment
that we considered erroneous, we always tried to also mention
references containing what we consider as the corresponding
correct description, or at least a right step towards that.
At first, it might appear that we claim that
theorists are usually correct while experimentalists are
prone to errors. This does definitely not reflect our view
of the field: papers that were criticised here for some
reason, still contain otherwise excellent work. For example,
a paper with whose mode-volume estimation we disagreed
\cite{zengin_prl114} is nevertheless a pioneering case of
designing plasmon--exciton hybrids operating at room temperature,
and also discusses thoroughly the criteria for reaching
strong coupling~\cite{zengin_jpcc120}. Our feeling is that
questionable extension of ideas has occurred mostly as a
result of pressure to differentiate new manuscripts by adding an
extra flavour. In the absence of a robust theoretical description,
approximations have therefore been made based on the theories
already available. This is of course acceptable, as long as
no attempt to interpret such estimates as exact quantified
results is made.

In summary, we have discussed situations where concepts from
cQED have been used to describe strong coupling in nanophotonics
--- and plasmonics in particular ---
without this use being fully justified. We showed that while
basing qualitative discussions on the factor $N/V$ can be a
good starting point,
as long as this factor is interpreted as an effective density of emitters,
in most prevalent architectures in current literature neither
$V$ nor $N$ are well defined, and quantitative conclusions
uniquely based on these quantities should be avoided, especially when excitons in
TMDs are involved. Instead, one should go one step backwards,
deduce experimentally relevant information directly from the
coupling of the standard dipole approximation, and then
measure if the system exhibits the desired quantum response.
We hope that the discussion presented here, even if it appears overly
critical at points, will help the nanophotonics community
to set its future efforts on a more solid foundation.

%
\begin{acknowledgements}
N.~A.~M. is a VILLUM Investigator supported by
VILLUM FONDEN (grant No. 16498).
S.~I.~B. acknowledges the European Research Council
(grant 341054, PLAQNAP).
The Center for Nano Optics is financially supported
by the University of Southern Denmark (SDU 2020 funding).
The Center for Nanostructured Graphene is sponsored by
the Danish National Research Foundation (Project No. DNRF103).
F.~T. and C.~W. acknowledge funding from MULTIPLY fellowships
under the Marie Sk\l{}odowska-Curie COFUND Action (grant
agreement No. 713694).
K.~B. acknowledges funding from the Deutsche
Forschungsgemeinschaft (DFG-Projektnummer
182087777-SFB 951 "HIOS").
We are grateful to L.~Lavazza for stimulating the work
and we thank T.~Shegai and J.~Aizpurua,
and the anonymous referees
for highly constructive feedback on the manuscript and
inspiring discussions.
\end{acknowledgements}

\end{document}